\documentclass{sf2a-conf2015}
\usepackage[english]{babel}
\usepackage{graphicx}
\usepackage[]{natbib}  
\usepackage{epstopdf}
\usepackage{hyperref} 

\def\BibTeX{{\rm B\kern-.05em{\sc i\kern-.025em b}\kern-.08em
    T\kern-.1667em\lower.7ex\hbox{E}\kern-.125emX}}
\bibpunct{(}{)}{;}{a}{}{,}  %%%%%%%%%%%%%  A&A bibliography style
\begin{document}

\TitreGlobal{SF2A 2015}

%%-----------------------------------------------------------------
%%      the top matter
%%

\title{Warm molecular Hydrogen at high redshift with the James Webb Space Telescope}

\runningtitle{Warm molecular Hydrogen at high redshift with JWST}

\author{P. Guillard}\address{Sorbonne Universit\'es, UPMC Universit\'e Paris 6 et CNRS, UMR 7095,  Institut d'Astrophysique de Paris, 98 bis bd Arago, 75014 Paris, France.}

\author{F. Boulanger}\address{Institut d'Astrophysique Spatiale, UMR 8617, CNRS, Universit\'e Paris-Sud, Bat. 121, 91405 Orsay, France}

%% IF Author3 has the same affiliation than Author1:
\author{M. D. Lehnert$^1$}

\author{P. N. Appleton}\address{NASA Herschel Science Center, Infrared Processing \& Analysis Center, California Institute of Technology, Pasadena, CA 91125, USA}

\author{G. Pineau des For\^ets$^{2,}$}\address{Observatoire de Paris, LERMA, UMR 8112, CNRS, 61 Avenue de l'Observatoire, 75014 Paris, France}

%% Keep this line, even if the page will be settled afterwards.
\setcounter{page}{237}

%%-----------------------------------------------------------------

\maketitle

%%-----------------------------------------------------------------
%%        The abstract
%% 
%%  Warning!  within the abstract:
%%  - do not use macros. 
%%  - do not use commands like: \cite, \citet, \citep ... etc.

\begin{abstract}
The build-up of galaxies is regulated by a complex interplay between gravitational collapse, galaxy merging and feedback related to AGN and star formation. The energy released by these processes has to dissipate for gas to cool, condense, and form stars. How gas cools is thus a key to understand galaxy formation. \textit{Spitzer Space Telescope} infrared spectroscopy revealed a population of galaxies with weak star formation and unusually powerful H$_2$ line emission. This is a signature of  turbulent dissipation, sustained by large-scale mechanical energy injection. The cooling of the multiphase interstellar medium is associated with emission in the H$_2$ lines. These results have profound consequences on our understanding of regulation of star formation, feedback and energetics of galaxy formation in general. The fact that H$_2$ lines can be strongly enhanced in high-redshift turbulent galaxies will be of great importance for the \textit{James Webb Space Telescope} observations which will unveil the role that H$_2$ plays as a cooling agent in the era of galaxy assembly.
\end{abstract}

%% Insert the keywords (to appear in the ADS indexing)
%% Keywords must be separated by a comma

%%-----------------------------------------------------------------

\section{Introduction: gas heating and cooling in galaxy assembly}
%%---------------------

In the $\Lambda$CDM framework, galaxies are assembled from the collapse of gas in virialized dark matter haloes \citep[e.g][]{White1978, Fall1980}. The most outstanding question in all contemporary theoretical studies of galaxies evolution is what processes regulate the gas content of galaxies, that is, the balance between accretion and mass loss. This balance, and thus the build-up of baryonic mass in galaxies, is regulated by a complex interplay between gravitational collapse, gas accretion, galaxy merging and feedback related to active galactic nuclei (AGN) activity and star formation \citep[e.g.][]{Dekel2006}. It is this competition between the rates of inflow, outflow, and star formation that gives the properties and physical characteristics of the galaxies we observe today. What currently limits our understanding of galaxy formation is how does the gas respond to those feedback mechanisms, which may inject sufficient mechanical energy into the interstellar medium (ISM) to have a major impact on star formation and galaxy assembly, thus potentially regulating the growth of galaxies \citep{Lehnert2015, Guillard2015}. Those feedback processes  will be particularly important during the early phases of galaxy
evolution at high redshift, when galaxies were gas-rich and most of the stars in the universe were
formed. 

The energy injected by feedback processes has to be dissipated for gas to cool and form stars. Observations of galaxies experiencing strong feedback and turbulence (e.g. galaxy interactions, AGN, cluster cooling flows) show that a significant fraction of this energy cascades down to small scales and is dissipated through line emission. This turbulent cascade is associated with the formation of multiphase ISM and one of the dominant cooling channel is through H$_2$ line emission \citep{Guillard2009, Guillard2012}. H$_2$ is a natural outcome of gas cooling, and the material from which stars are formed. Being affected by star formation, massive central black holes, and inflows and outflows of gas, H$_2$ plays an important role in all stages of galaxy formation and evolution \citep{Boulanger2009}. 
% is also the seed of a network of chemical reactions, ultimately leading to molecular complexity in space \citep[e.g.][]{Herbst1995}.
In this paper, we stress the unique capability of the James Web Space Telescope to detect and characterize H$_2$ line emission at the peak of the star-forming activity of the Universe. Those observations will be key to study the structure and phase distribution of the gas, because they will allow us to estimate the cooling, turbulent dissipation, and dynamical times.
  
\section{H$_2$ line emission as a probe of the energetics of molecular gas}
%%--------------------------------------------
% FIGURES CRITICAL DENSITIES
%----------------------------------------------
\begin{figure}[t]
 \centering
 \includegraphics[width=0.49\textwidth,clip]{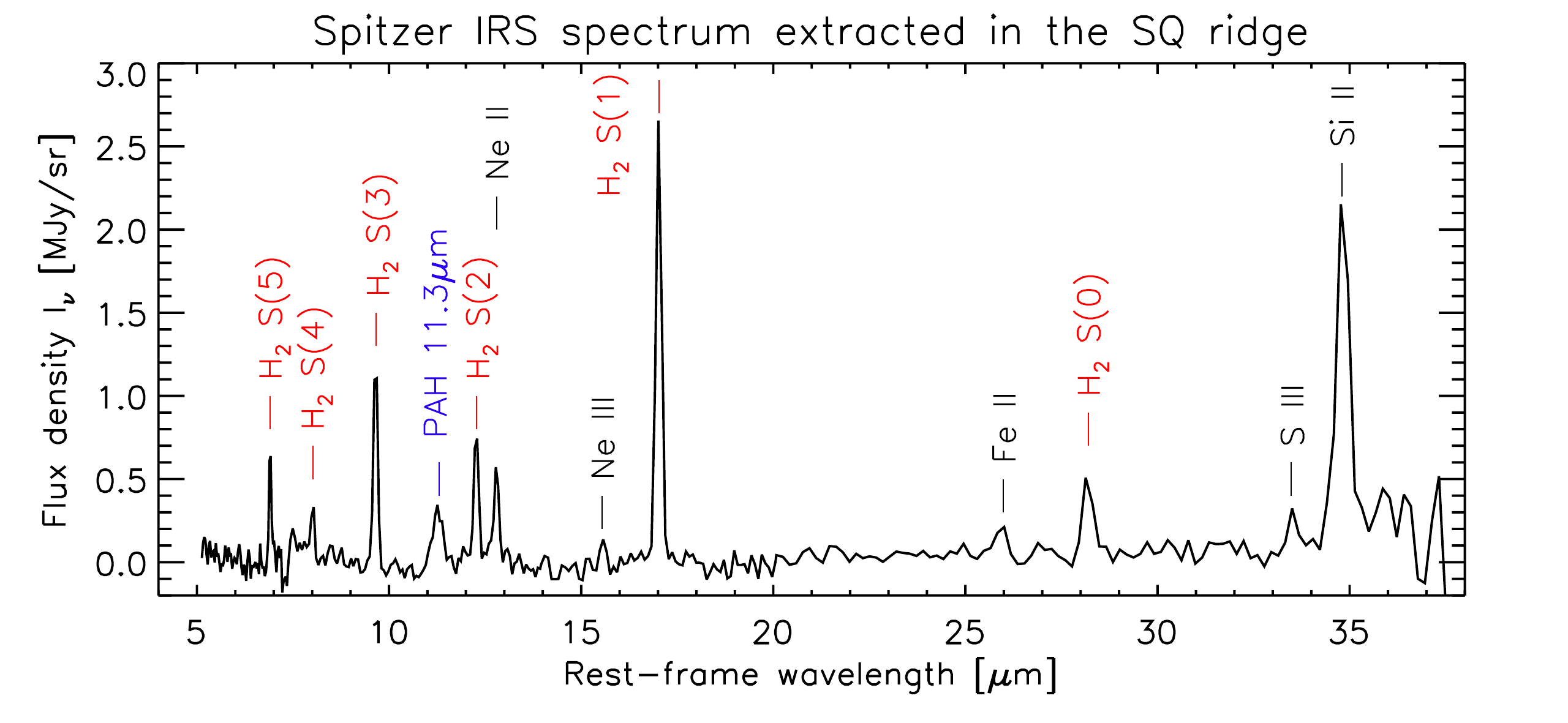}      
  \includegraphics[width=0.49\textwidth,clip]{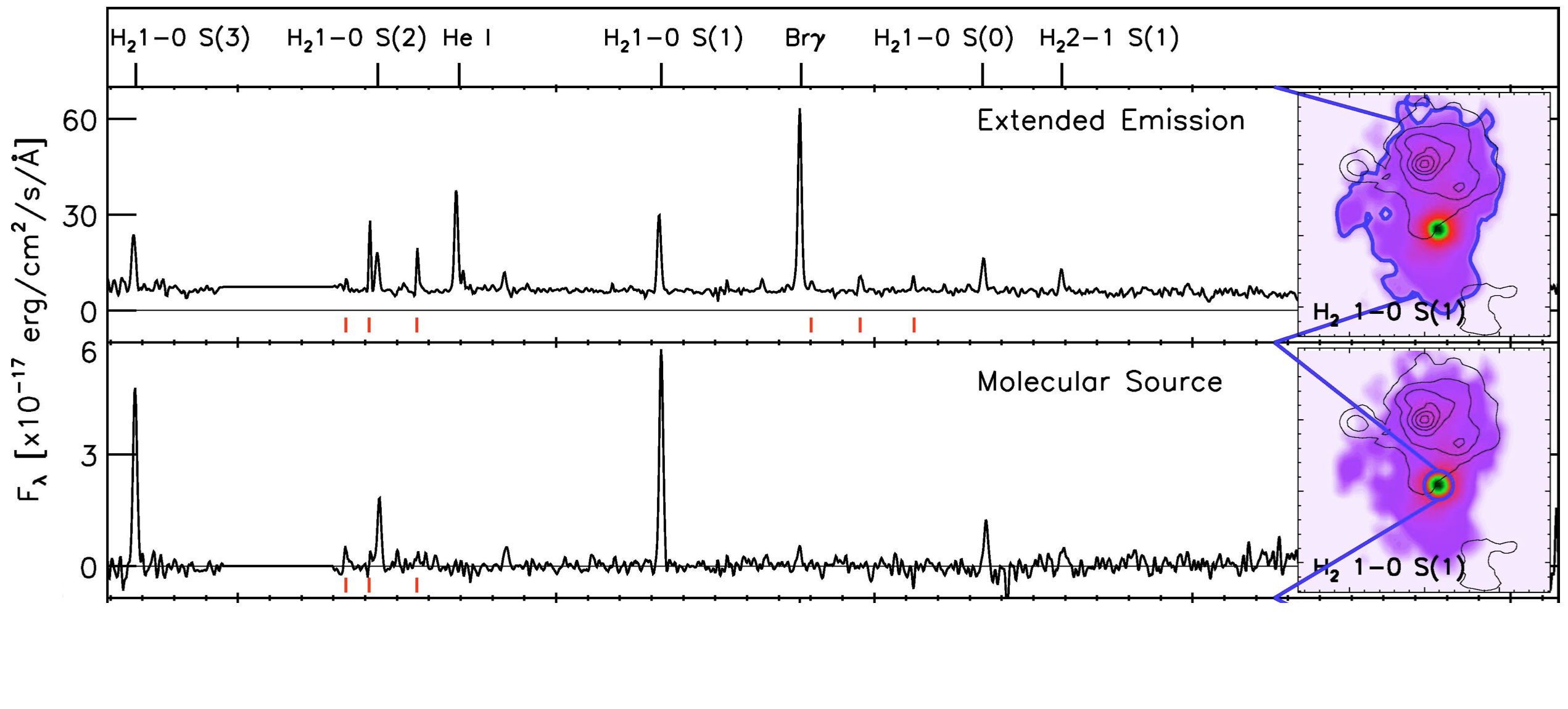} 
  \caption{\textit{Left:} Spitzer IRS mid-infrared spectrum of the Stephan's Quintet intra-group medium \citep{Appleton2006, Cluver2010}. \textit{Right:} Integrated VLT/SINFONI near-infrared spectra of two regions (a compact and extended one) in the overlap area of the Antennae galaxies \citep{Herrera2011}. In the right panel, blue contours mark the aperture from which each spectrum was extracted. In both sources, the spectra show prominent H$_2$ rotational and ro-vibrational lines, respectively. The H$_2$ emission is originating from the dissipation of turbulent energy driven by large-scale gas dynamics, a galaxy hitting a tidal filament for Stephan's Quintet and the formation of bound clouds through accretion for the Antennae overlap region.}
  \label{fig1}
\end{figure}

Excitation of rotation-vibration levels of H$_2$ can occur through different mechanisms. Collisional excitation with atoms and molecules \citep[e.g.][]{Flower1998, LeBourlot2002}, absorption of UV photons followed by fluorescence \citep[e.g.][]{Gautier1976, Black1987}, heating by hard X-rays penetrating into the molecular clouds \citep{Maloney1996, Tine1997}, and  cosmic ray heating \citep{Dalgarno1999}, which has mainly been discussed for the strong H$_2$ emission in cooling-flow filaments \citep{Ferland2008}.

Rotational and ro-vibrational lines of molecular hydrogen (H$_2$) have become an important diagnostic tool for shocks in the galactic \citep[e.g.][]{Allen1993, Falgarone2005, Hewitt2009, Ingalls2011} and extra-galactic interstellar medium \citep[e.g.][]{Wright1993, Appleton2006, Veilleux2009, Ogle2010, Beirao2015}. Two examples are given in Figure~\ref{fig1}. Pure rotational lines of H$_2$ (0-0 S(0), 0-0 S(1), etc.) are found in the mid-infrared between $3-30\,\mu$m and trace warm gas with typical temperatures of a few 100 K to 1000 K. Ro-vibrational lines of H$_2$ (e.g., 1-0 S(1) at 2.12$\,\mu$m) are observed in the near-infrared and trace hotter gas with temperatures of a few 1000 K. According to shock models, H$_2$ line ratios can be used to infer the pre-shock gas characteristics (density, magnetic field) and shock properties (velocity, non-dissociative -- C-type -- or dissociative -- J-type --) \citep{Flower2010, Guillard2009, Guillard2012}.

\section{Tracing the kinetic energy dissipation associated with feedback processes at $\bf z \sim 1.5 - 3.5$ with JWST}

\begin{figure}[ht!]
 \centering
 \includegraphics[width=0.93\textwidth]{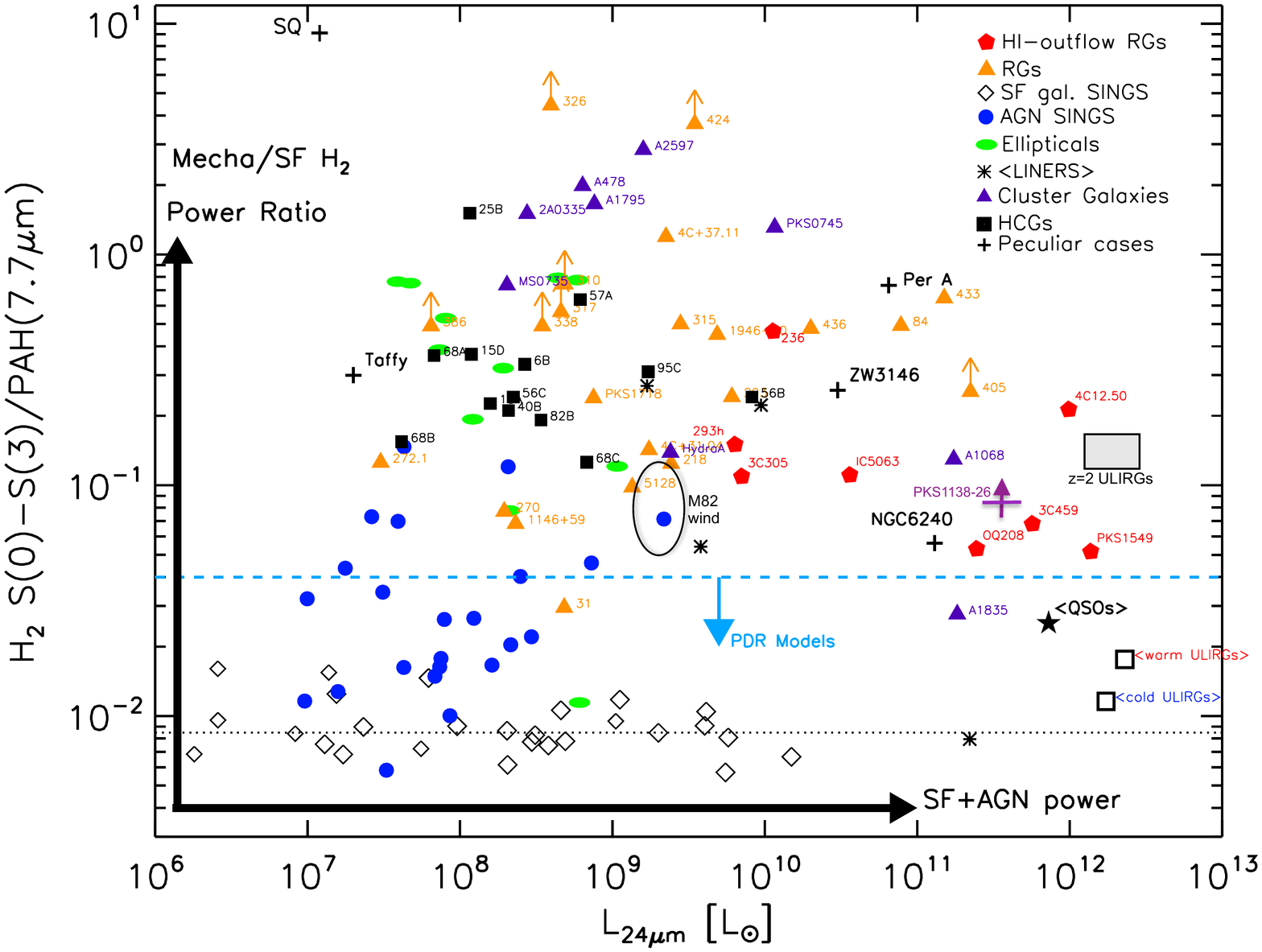}      
  \caption{Ratio of the mid-IR $\rm H_2$ line luminosities (summed over S(0) to S(3)) to the PAH 7.7$\mu$m emission vs. 24$\mu$m continuum luminosity \citep[updated from][]{Guillard2012}. This ratio indicates the relative contribution of mechanical heating (shocks) and star-formation (SF) power (UV excitation). The red pentagons are nearby radio galaxies with fast ($> \! 1000\,$km/s) H{\sc i} outflows observed with \textit{Spitzer IRS} \citep{Guillard2012}. The orange triangles and green ellipses are samples of radio galaxies \citep[respectively][]{Ogle2010, Kaneda2008}.  These $\rm H_2$-luminous galaxies stand out above SF and AGN galaxies from the SINGS survey \citep{Roussel2007}. The $\rm H_2$ emission in these sources cannot be accounted by UV or X-ray photon heating. The blue dashed line shows the upper limit given by the \citet{Kaufman2006} PDR models ($n_{\rm H} = 10^4\,$cm$^{-3}$, $G_{\rm UV}=10$). For comparison, a few other types of $\rm H_2$-luminous galaxies are shown: the Stephan's Quintet (SQ) and Taffy galaxy collisions \citep{Cluver2010, Peterson2012}, other Hickson Compact Groups \citep[black squares,][]{Cluver2013}, the ZW 3146 \citep{Egami2006} and Perseus A \citep{Johnstone2007} clusters, and the NGC 6240 merger \citep{Armus2006}. The black ellipse shows the \textit{Spitzer IRS} observations of the M82 wind \citep{Beirao2015}, the black rectangle shows the detection of H$_2$ in stacked \textit{Spitzer} spectra of $z=2$ ULIRGs \citep{Fiolet2010}, and the purple cross the detection of H$_2$ in the Spider Web (PKS1138-26) radio galaxy protocluster \citep{Ogle2012}.}
  \label{fig2}
\end{figure}

The \textit{James Webb Space Telescope} will be, 15 years after the \textit{Spitzer Space Telescope}, the next mission to have access to rotation-vibration H$_2$ transitions. As such, it will play a critical role in the context of galaxy evolution since H$_2$ represents an important, if not dominant, cooling agent in the energetics of galaxy formation.
Observations by the InfraRed Spectrograph (IRS) onboard the  {\it Spitzer Space Telescope} unveiled a significant and diverse population of low-$z$ objects where the
mid-infrared rotational line emission of H$_2$ is strongly enhanced ($L_{\rm H_2} \sim 10^{40} - 10^{44}$~erg~s$^{-1}$), while star formation is suppressed (see Figure~\ref{fig2}). This suggest that shocks are the primary cause of the H$_2$ emission \citep{Guillard2009}. This sample of {\it H$_2$-luminous sources} includes galaxies in several key phases of their evolution, dominated by, for instance, gas accretion onto bright central galaxies in clusters \citep{Egami2006}, galaxy interactions \citep{Appleton2006}, or galactic winds driven by star formation \citep[e.g. M82][]{Beirao2015}, and radio-loud AGN \citep{Ogle2010, Guillard2012}. In those sources, the turbulent dissipation time is longer than the dynamical time, the mechanical energy contained in the molecular phase being dominant over the thermal energy of the gas \citep[e.g.][]{Guillard2012a}.

Constraining the impact of merging and AGN feedback on the formation and evolution of massive galaxies
can only be addressed through direct H$_2$ line observations at $z \sim 2$, near the cosmologically most active period of star formation, galaxy interactions and AGN activity. By analogy to what is observed on local H$_2$-luminous objects, we expect the mid-IR lines to be the dominant cooling lines for warm, $10^{2-3}$~K, gas in the strongly shocked, highly turbulent, colliding flows in galaxy interactions \citep[e.g. the galaxy-wide shock in Stephan's Quintet,][]{Guillard2009}, but also, e.g., in AGN-driven outflows.
High gas velocity dispersions measured in $z \sim 2$ actively star-forming galaxies show that the gas kinematics in these systems was strongly disturbed compared to galaxies today \citep[e.g.][]{Lehnert2009}. The molecular gas is observed to be highly turbulent and therefore the warm H$_2$ emission is expected to be more frequent and more powerful than at low-$z$, as suggested by H$_2$ detections in $z \approx 2$ infrared-luminous galaxies \citep{Fiolet2010, Ogle2012}. In those sources, H$_2$ line emission is likely powered by the dissipation of turbulence, which could originate from star formation (supernovae), radiation pressure, or gas accretion.

%%%%%%%%%%%%%%%%%%%%%%%%%%%%%%%%%%%%%%%%
%		FIGURE: OBSERVING H$2WITHIN MIRI / MRS
%%%%%%%%%%%%%%%%%%%%%%%%%%%%%%%%%%%%%%%%
\begin{figure}
   \centering
    \includegraphics[angle=90, width=0.65\textwidth]{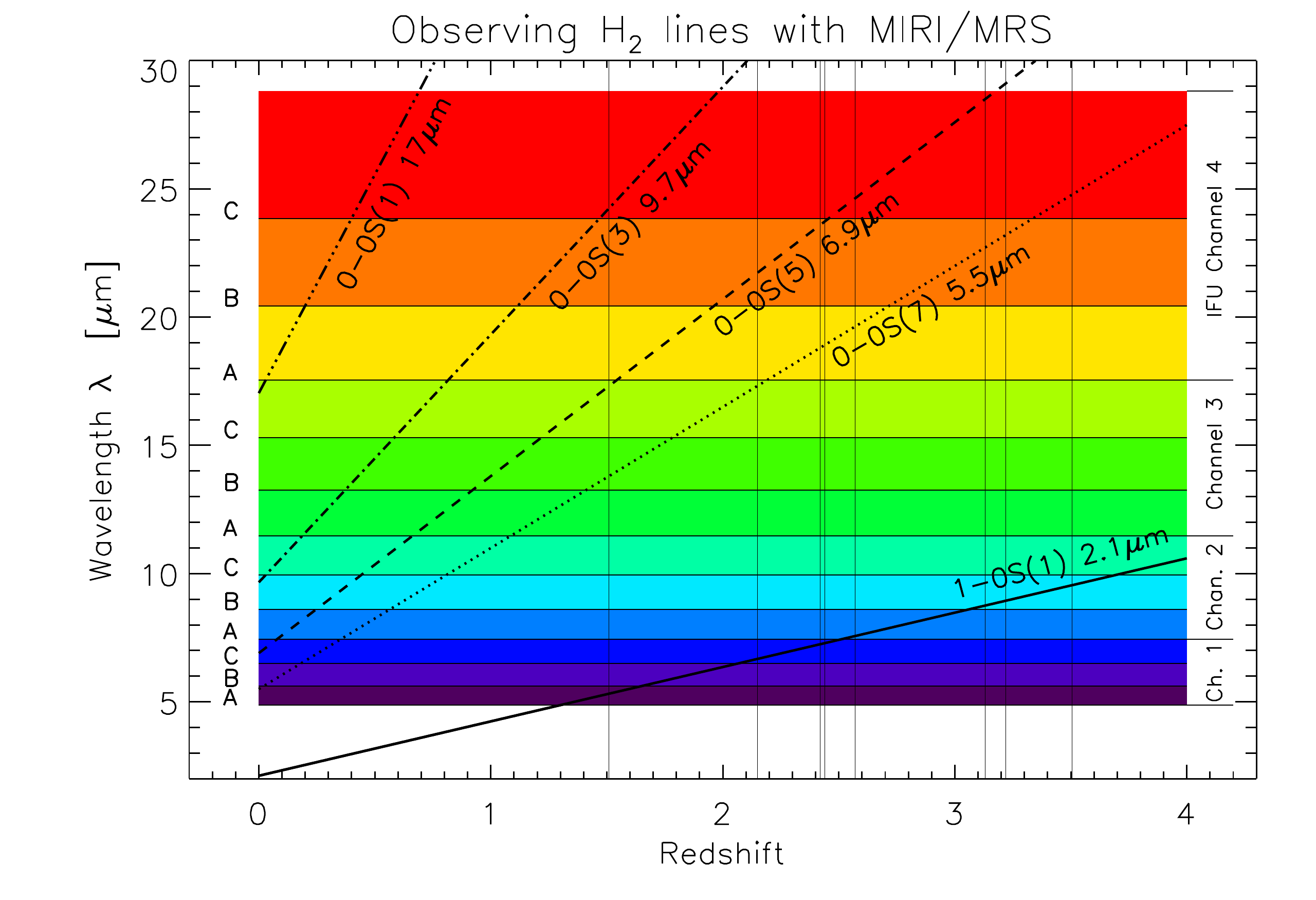}
      \caption[Observing H$_2$ lines at high-redshift with JWST/MIRI]{Observing H$_2$ lines at high-redshift with JWST/MIRI. The observed wavelengths of some H$_2$ lines are shown as a function of the redshift. The colored bars indicate the channels and bands of the Medium Resolution Spectrometer (MRS) of the MIRI instrument. One observation corresponds to four sub-bands, like 1A, 2A, 3A, 4A for instance \citep[see][for technical details about the MRS operations]{Guillard2010a}. The vertical lines indicate the redshifts of some high-$z$ radio-galaxies that might be interesting to look at with MIRI.}
       \label{fig3}
\end{figure}

Covering a wavelength range of $4.9 - 28.6\, \mu$m, the MIRI Medium Resolution Spectrometer \citep[MRS,][]{Wells2015} will be the first Integral Field Unit (IFU) instrument to provide the sensitivity and resolving power to spatially and spectrally resolve H$_2$ and forbidden ionized gas lines at rest-frame near-IR and mid-IR wavelengths, out to $z = 1.5 - 3.5$ (Figure~\ref{fig3}). 
Covering $0.6 - 5\, \mu$m in the near-infrared, NIRSPEC \citep{Posselt2004} will allow the detection of ro-vibrational lines at very high sensitivity ($0.6 \times 10^{-21}$~W~m$^{-2}$ for 1h) and spectral resolution ($R \approx 3000$ for the IFU mode). Both the MRS and NIRSPEC IFUs will have comparable spectral resolutions to the SINFONI near-infrared IFU on the VLT (see Figure~\ref{fig1}, right panel). The JWST instruments will allow us to directly investigate  the physical state and the kinematics of the ionized gas and the warm ($> 150$~K) molecular gas that is dynamically heated by the dissipation of mechanical energy associated with galaxy merging and AGN feedback. To establish the energy budget of the warm molecular gas and shock diagnostics, we shall use the near-IR ro-vibrational lines, e.g. the H$_2$ 1-0 S(1) $2.12 \, \mu$m line, and the mid-IR pure rotational H$_2$ 0-0 S(3) $9.7 \, \mu$m and S(5) $6.9 \, \mu$m lines. The forbidden ionized gas lines (e.g. [Ne II], [Ne III]) will be used to compare the kinematics of the molecular gas with that of the ionized gas and complement the shock diagnostics \citep{Gusdorf2015}. The synergy with NIRSPEC will be helpful to observe the CO bandheads and Ca$\,${\sc II} triplet to estimate the stellar kinematics. This will allow to estimate the ratio between the bulk galaxy rotation and the gas velocity dispersion in these high-$z$ objects, and provide an absolute rest-frame in which to interpret the gas motions as blueshift or redshift.

\section{Conclusions and perspectives}

Rotation-vibration H$_2$ transitions observed in the mid- and near-infrared appear as key tracers of the energetics of galaxy formation and evolution. They complement the CO transitions which usually trace the bulk of the molecular gas that is too cold to emit in H$_2$. The powerful infrared H$_2$ line emission observed in a large sample of extragalactic sources is believed to be powered by the dissipation of turbulent energy, provided by large-scale shocks from galaxy collisions, radio jet feedback, star formation and gas accretion. In some cases, the line emission represents a significant fraction of the total molecular gas mass and bolometric luminosity of the galaxies. By observing routinely those lines, the JWST should allow us to relate the star formation activity and the gas accretion rates to the turbulence of the gas. This has potentially far-reaching implications, from the physics of the multiphase ISM, regulation of star formation in the most massive galaxies, and the formation of the first galaxies.

% Optional acknowledgements
% -------------------------
%\begin{acknowledgements}
%We thank...
%\end{acknowledgements}
%

%% The following lines are required when using BibTEX (strongly encouraged!):
\bibliographystyle{aa}  % A&A bibliography style file (aa.bst)
\bibliography{guillard.bbl} % your references in file: Yourfile.bib
%\bibliography{/Users/guillard/bibliography/biblio_bibtex/biblio.bib}

%
\end{document}